\documentclass[prl,aps,twocolumn,showpacs,superscriptaddress,floatfix,amsmath,amssymb]{revtex4}

\usepackage{graphicx}% Include figure files
\usepackage{dcolumn}% Align table columns on decimal point
\usepackage{bm}% bold math
\usepackage[dvipdfm,bookmarks=true,colorlinks,%
            citecolor=blue,linkcolor=blue, hypertex, %
            breaklinks=true]{hyperref}

\usepackage{ulem}

\def\be{\begin{eqnarray}}
\def\ee{\end{eqnarray}}

\newcommand{\Fig}[1]{Fig.~\ref{#1}}

\begin{document}

\title{Partial order in Potts models on the generalized decorated square
lattice}

\author{ M. P. Qin}
%\email {qiaoni2233@gmail.com}
\affiliation{Institute of Physics, Chinese Academy of Sciences, Beijing
100190, China}

\author{J. Chen}
\affiliation{Institute of Physics, Chinese Academy of Sciences, Beijing
100190, China}

\author{Q. N. Chen}
\affiliation{Department of Chemistry, Frick Laboratory, Princeton University,
Princeton, New Jersey 08544, USA}

\author{Z. Y. Xie}
\affiliation{Institute of Physics, Chinese Academy of Sciences, Beijing
100190, China}

\author{X. Kong}
\affiliation{Institute of Physics, Chinese Academy of Sciences, Beijing
100190, China}

\author{H. H. Zhao}
\affiliation{Institute of Physics, Chinese Academy of Sciences, Beijing
100190, China}

\author{B. Normand}
\affiliation{Department of Physics, Renmin University of China, Beijing
100872, China}

\author{T. Xiang}
%\email {txiang@aphy.iphy.ac.cn}
\affiliation{Institute of Physics, Chinese Academy of Sciences, Beijing
100190, China}

\date{\today}

\begin{abstract}
We explore the Potts model on the generalized decorated square lattice, with
both nearest ($J_{1}$) and next-neighbor ($J_{2}$) interactions. Using
the tensor renormalization-group method augmented by higher-order singular
value decompositions, we calculate the spontaneous magnetization of the
Potts model with $q = 2$, 3, and 4. The results for $q = 2$ allow us to
benchmark our numerics using the exact solution. For $q = 3$, we find a
highly degenerate ground state with partial order on a single sublattice,
but with vanishing entropy per site, and we obtain the phase diagram as a
function of the ratio $J_{2}/J_{1}$. There is no finite-temperature transition
for the $q = 4$ case when $J_{1} = J_{2}$, but the magnetic susceptibility
diverges as the temperature goes to zero, showing that the model is critical
at $T = 0$.
\end{abstract}


\maketitle

The $q$-state Potts model plays an important role in the understanding of
different phases and critical phenomena (for a review see Ref.~\cite{wu_1}).
The behavior of ferromagnetic Potts models is well understood due to their
universality. The properties of antiferromagnetic Potts models are much more
complex, varying widely for different values of $q$ and lattice topologies.
On every lattice there exists a value $q_{c}$ such that for $q > q_{c}$ the
model has exponentially decaying correlations and hence no order at all
temperatures, including $T = 0$. For $q = q_{c}$, the model has a critical
point at zero temperature, and for $q < q_{c}$ all types of behavior are
possible \cite{sokal_qc}, including long-ranged order or partial order.
When the lattice is irregular, meaning that not all sites are equivalent,
entropy-driven phase transtions and exotic forms of partial order may arise
\cite{Sokal_diced,qiaoni,youjin}. Partial order has been found both in
classical statistical models \cite{Diep_1,Lipowski} and in quantum systems
\cite{Diep_2}.

\begin{figure}[tbp]
\begin{center}
\includegraphics[scale=0.4]{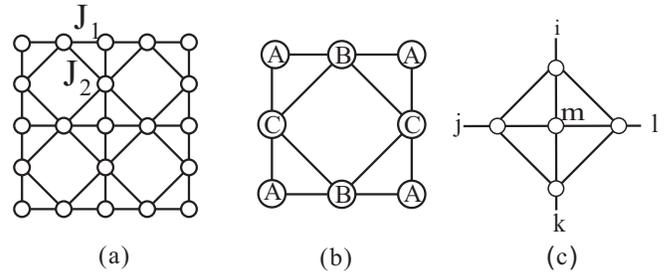}
\caption{(a) Generalized decorated square lattice with interactions between
decorating sites. (b) Sublattice labels for this tripartite lattice. (c)
Label definition for the tensor expressed in Eq.~(\ref{T}).}
\label{lattice}
\end{center}
\end{figure}

The decorated square lattice is formed by adding one site to each bond of a
square lattice. Here we study Potts models on the generalized decorated square
lattice obtained by introducing an interaction between the decorating sites
[\Fig{lattice}(a)]. This is a tripartite lattice, for which one attribution
of sublattices ($A,B,C$) is shown in \Fig{lattice}(b). The Hamiltonian of
the model is
\begin{equation}
H = J_{1} \sum\limits_{\langle i,j \rangle} \delta_{\sigma_i,\sigma_j} + J_{2}
\sum\limits_{\langle\langle i,j \rangle\rangle} \delta_{\sigma_i,\sigma_j} - h
\sum\limits_{i \in \cal{L}}
\delta_{\sigma_i, 0},
\label{ham}
\end{equation}
where $\sigma _{i} = 0, 1, 2, \dots $ is the Potts state on lattice site
$i$. The first term denotes all nearest-neighbor bonds and the second all
next-neighbor bonds in \Fig{lattice}(a), while the third term denotes a
field term $h$ applied to the sites in any sublattice(s) $\cal{L}$ for
the calculation of magnetizations and susceptibilities.

In the pure decorated square lattice, with only nearest-neighbor ($J_1$)
interactions, taking a partial trace over the decorating ($B$ and $C$) sites
effects a mapping to a ferromagnetic Potts model on the square lattice
\cite{wu_2,Yosio,Hajdukovic}. From this one may deduce that the critical
value for the $q$-state antiferromagnetic Potts model on the decorated
square lattice is $q_{c} = \frac{1}{2}(3 + \sqrt{5}) = 2.618$, meaning
that for $q = 3$ and $q = 4$ there is no ordered phase at low temperatures
and no phase transition occurs. This value of $q_c$ is smaller than that for
the antiferromagnetic Potts model on the square lattice, where $q_c = 3$
\cite{Nightingale,den,Racz,Reed,sokal_q_3}. For $q = 2$ (the Ising model),
the exact solution is one with complete long-range order \cite{Diep_4}.

When next-neighbor interactions are introduced, reentrant behavior of the
ordered phases has been found both for $q = 2$ \cite{Diep_3,Diep_4} and
$q = 3$ \cite{su}. However, this phenomenon required that one of the
interactions be ferromagnetic \cite{su}. Here we consider the regime
where both $J_1$ and $J_2$ are antiferromagnetic and focus primarily on
the isotropic situation ($J_1 = J_2$), where the competition between the
different interactions is the largest.

For a classical model on any lattice, it is always possible to find a
tensor-network representation of the partition function \cite{levin,huihai}.
For the Hamiltonian defined in Eq.~(\ref{ham}), we define a rank-four
tensor $T_{ijkl}$ for the unit cell shown in \Fig{lattice}(c) as
\begin{eqnarray}
T_{ijkl} = \sum \limits_{m} \!\!\! & \exp & \!\!\! [ - \beta J_{1}
(\delta_{\sigma_i,\sigma_m} \! + \delta_{\sigma_j,\sigma_m} \! + \delta_{\sigma_k,\sigma_m}
\! + \delta_{\sigma_l,\sigma_m} ) ] \! \times \nonumber \\
&\exp& \!\!\! [ - \beta J_{2} ( \delta_{\sigma_i,\sigma_j} \!
 + \delta_{\sigma_j,\sigma_k} \! + \delta_{\sigma_k,\sigma_l} \!
 + \delta_{\sigma_l,\sigma_i} ) ] .
\label{T}
\end{eqnarray}
The Potts variables $\sigma_i$ on each site serve as the indices of the
tensors, and the tensors for each unit cell form a square-lattice tensor
network.

Recently, we have developed a Tensor Renormalization Group (TRG) technique
which makes use of higher-order singular value decomposition (HOSVD) of
the tensors to truncate them during the renormalization (coarse-graining)
process \cite{zhiyuan_ho}. There are two steps in this method at each
iteration $n$. In the first, we block two tensors to form a larger tensor,
\begin{equation}
M^{(n)}_{xx'yy'} = \sum_{i} T^{(n)}_{x_{1}x'_{1}yi} T^{(n)}_{x_{2}x'_{2}iy'} ,
\label{eq:2DFatTensor}
\end{equation}
where $x = x_1\otimes x_2$ and $x' = x'_1 \otimes x'_2$. In the second, we
perform a HOSVD,
\begin{equation}
M^{(n)}_{xx'yy'} = \sum_{ijkl} S_{ijkl} U^L_{xi} U^R_{x'j} U^U_{yk} U^D_{y'l} ,
\label{eq:2DHOSVD}
\end{equation}
where the $U$ matrices are unitary and $S$ is the core tensor of $M^{(n)}$,
which is fully orthogonal and pseudo-diagonal \cite{zhiyuan_ho}. The tensor
is then truncated according to
\begin{eqnarray}
T^{(n+1)}_{xx'yy'} = \sum_{ij} U^{(n+1)}_{ix} M^{(n)}_{ijyy'} U^{(n+1)}_{jx'}
\label{eq:2DTruncation}
\end{eqnarray}
to obtain a new tensor $T^{(n+1)}$ with the same dimension as $T^{(n)}$. In one
``HOTRG'' step, the size of the lattice is reduced by a factor of $1/2$. By
repeating the HOTRG step alternately in the $x$ and $y$ directions, we
contract the tensor network and obtain the partition function of the model.

While several methods exist to contract the
tensor network \cite{levin,huihai, vidal_1, vidal_2}, we have demonstrated
\cite{zhiyuan_ho} that the accuracy of this HOTRG method is very high:
a tensor dimension of just $D = 24$ in the two-dimensional Ising model is
sufficient to ensure that the relative error in the free energy is less
than $10^{-7}$, even at the critical temperature. The accuracy can be further
improved by including the effects of the environment in the renormalization
process, known as the second renormalization group (SRG) method
\cite{zhiyuan_srg}, although we do not need to implement this for the
accuracies attained here.

For numerical reasons, the thermodynamic quantities obtained most accurately
from the HOTRG partition function are the magnetization and susceptibility.
The small field term included in Eq.~\eqref{ham} pins a state of $q = 0$
order to a specific sublattice ${\cal L}$. For a starting square lattice of
dimensions $L$$\times$$L = N$, we define the order parameter as
\begin{equation}
M = \frac{1}{N} \langle \sum\limits_{i \in {\cal L}} \delta_{\sigma_i,0} \rangle
 - \frac{1}{q},
\label{mag_d}
\end{equation}
which is the expectation value of the number of spins in sublattice
${\cal L}$ whose value is $0$. The term $1/q$ is subtracted to ensure a
value of $0$ in the disordered phase. The susceptibility is given by $\chi
 = \partial M / \partial h$. The order parameter $M$ expresses the broken
sublattice symmetry \cite{Grest} of the partially ordered phase: only sites
of one sublattice are ordered, while all others are disordered (but may still
be correlated locally). An ordered state with broken sublattice symmetry was
first suggested for the $q = 3$ antiferromagnetic Potts model on the square
lattice, but more detailed analysis demonstrated that this order is
artificial, the model being disordered at all temperatures other than
a $T_c = 0$ critical point \cite{Nightingale,den,Racz,Reed}.

\begin{figure}[tbp]
\begin{center}
\includegraphics[scale=0.4]{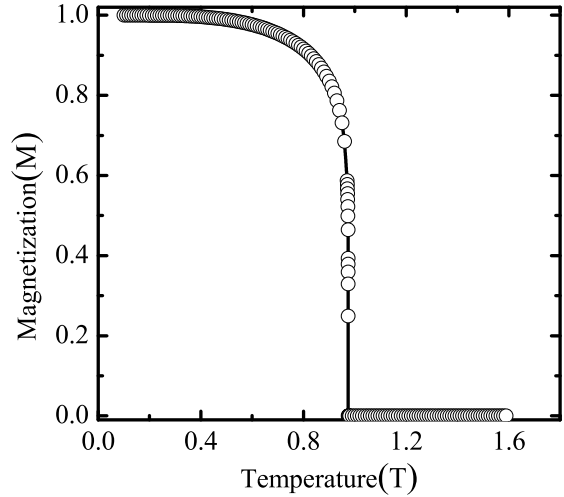}
\caption{Magnetization of the $q = 2$ Potts model on the generalized
decorated square lattice with $J_1 = J_2 = 1$. The bond dimension in the
HOTRG calculation is $D = 30$. An Ising-like phase transition occurs at
$T_{c} = 0.97423$}
\label{Mag_q_2}
\end{center}
\end{figure}

The generalized decorated square lattice model is exactly solvable
\cite{Diep_3} for $q = 2$. There are three sublattices but only two
possible states for each site, leading to frustration. Because the
coordination number of the $A$ sites, $z_A = 4$, is less than that of
the $B$ and $C$ sites ($z_B = z_C = 6$), the ground state is one with
N\'eel order on the $B$ and $C$ sublattices while the Ising spins on
the $A$ sites are free. This configuration minimizes the number of
unsatisfied bonds and hence the energy. The free spins are responsible
for the zero-temperature residual entropy, $S_{0} = \frac{1}{3} \ln{2}$
per site. The universality class of the phase transition from the N\'eel
phase to the paramagnetic phase remains that of the two-dimensional Ising
model. In \Fig{Mag_q_2} we show the magnetization obtained for $J_1 = J_2$
with tensor dimension $D = 30$ in the HOTRG calculation. The transition
temperature we compute is $T_{c} = 0.97423$, which is very close to the
exact result $T_c = 0.9742197$, demonstrating clearly the accuracy of
the HOTRG method.

\begin{figure}[tbp]
\begin{center}
\includegraphics[scale=0.11]{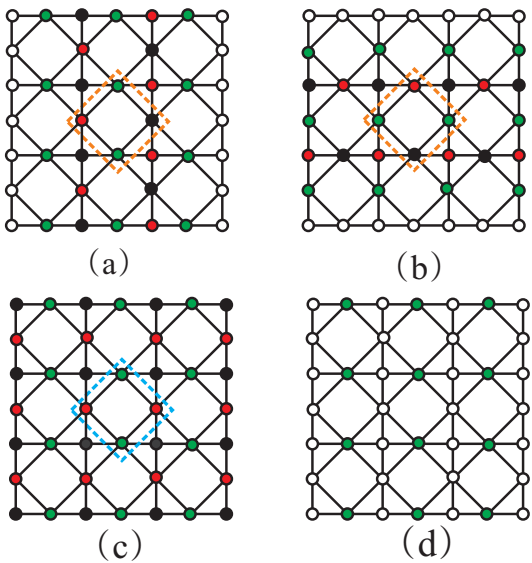}
\caption{Ground-state configurations for the $q = 3$ Potts model on the
generalized decorated square lattice. The three states are represented by
red for 0, black for 1, and green for 2. Panels (a) and (b) illustrate an
arbitrarily chosen square on the $B$ and $C$ sublattices with one pair of
diagonally opposite sites in the same state, which creates a complete
partial order on the corresponding sublattice. In panel (c), both opposite
pairs have the same state and all three sublattice are ordered. All of the
ground-state configurations are summarized in panel (d) and a $\pi/2$
rotation of this.}
\label{seed_all_new}
\end{center}
\end{figure}

We turn now to the $q = 3$ model and begin by developing a heuristic
understanding of the underlying physics. Because each site may take one of
three states and the lattice is tripartite, all bonds can be satisfied and
this model is unfrustrated. If the three states are represented by three
different colors, as in \Fig{seed_all_new}, the problem of finding all the
ground-state configurations may be cast as a coloring problem. In fact the
division of sites into three sublattices shown in \Fig{lattice}(b) is not
unique (a situation that would ensure the same type of order as in the
$q = 3$ Potts model on the triangular lattice). The origin of the different
possible sublattice designations is the fact that $z_A = 4$ for one of the
sublattices, introducing more degrees of freedom and causing a degenerate
ground state.

For the square of $B$- and $C$-sublattice sites marked by the orange dashed
line in Figs.~\ref{seed_all_new}(a) and (b), at least one pair of diagonally
opposite sites must have the same color. Each such square effectively
nucleates a state of complete order on the relevant sublattice, as
shown in Figs.~\ref{seed_all_new}(a) and (b), without ordering the other
two sublattices (each of which in fact forms a state of two-fold degenerate
alternating lines). If both pairs of opposite sites have the same color
[\Fig{seed_all_new}(c)], then the nucleated state would be one of complete
order on all three sublattices, but this is a set of measure zero. The set
of all ground-state configurations is summarized in \Fig{seed_all_new}(d),
and it is one with partial order on a single sublattice. Thus the additional
freedom contained in the special topology of the generalized decorated square
lattice is not sufficient to drive the system into a disordered phase.

To investigate the nature of the partially ordered state, if the $B$
sublattice is ordered with $q = 0$, the state of the intervening lines
of $A$ and $C$ sites may be either $1212 \dots 12$ or $2121 \dots 21$
at random, and similarly for an ordered $C$ sublattice. Thus the partially
ordered state breaks the $\pi/2$ rotation symmetry of the lattice. The
ground-state degeneracy at zero temperature is $N_{0} = 6 (2^{L} - 1)$,
where $6$ is subtracted because configurations in which both $B$ and $C$
sublattices are ordered [\Fig{seed_all_new}(c)] are double-counted. Despite
this high degneracy, the residual entropy per site,
\begin{equation}
S_{0} = \lim_{L \rightarrow \infty } \frac{\log (6\times 2^{L} - 6)}{L^{2}} = 0,
\end{equation}
vanishes because of the one-dimensional nature of the freedom in ground-state
configurations within a two-dimensional lattice. This type of partial order
is different from the order found in Refs.~\cite{qiaoni,Grest}, where only
one sublattice was ordered and the other sites remained in random states.
For the generalized decorated square lattice, the sites on lines between the
ordered sublattice sites remain highly correlated, forming a one-dimensional
antiferromagnetic Ising model ($q = 2$), which has two-fold degeneracy
between the alternating states as noted above.

\begin{figure}[tbp]
\begin{center}
\includegraphics[scale=0.4]{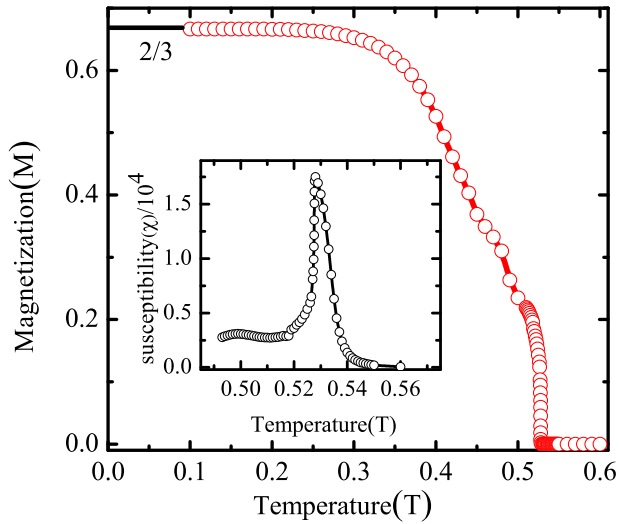}
\caption{Magnetization for the $q = 3$ Potts model on the generalized
decorated square lattice with $J_1 = J_2$ and $D = 30$. A second-order phase
transition occurs at $T_{c} = 0.528(1)$. The inset shows the susceptibility,
which diverges at $T_c$.}
\label{Mag_q-3}
\end{center}
\end{figure}

We calculate the magnetization and susceptibility for the $q = 3$ Potts
model on the generalized decorated square lattice for $J_1 = J_2$ using
the HOTRG method with $D = 30$. As shown in \Fig{Mag_q-3}, we find that
the magnetization goes continuously to zero at a second-order phase
transition at $T_{c} = 0.528(1)$. At low temperatures, the magnetization
defined by Eq.~(\ref{mag_d}) with ${\cal L} \equiv C$ takes the value
2/3, which means that all $C$ sites are fully polarized and confirms our
analysis concerning the nature of the ground state. The susceptibility
(inset, \Fig{Mag_q-3}) shows a clear divergence at $T_{c}$, confirming
the second-order nature of the phase transition.

\begin{figure}[tbp]
\begin{center}
\includegraphics[scale=0.4]{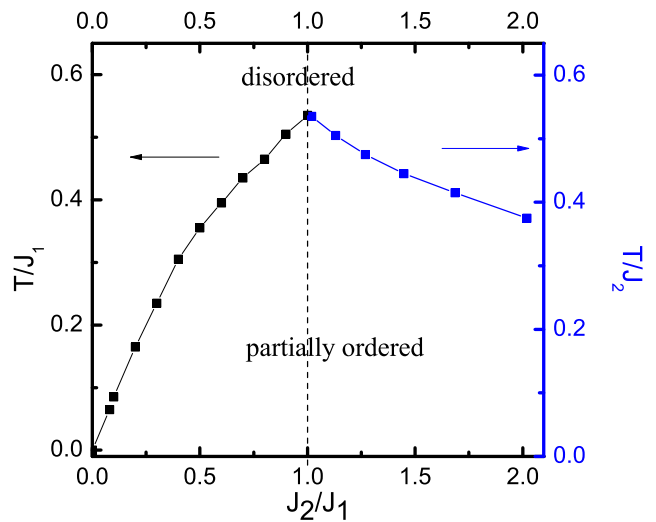}
\caption{Phase diagram for the $q = 3$ Potts model on the generalized
decorated square lattice as a function of $J_2/J_1$. HOTRG calculations
for the critical temperature $T_c$ were performed with $D = 30$. The
temperature is expressed in units of $J_1$ for $J_2/J_1 < 1$ and in units
of $J_2$ when $J_2/J_1 > 1$.}
\label{phase}
\end{center}
\end{figure}

Having established the physics of partial order where $J_1 = J_2$, we
next investigate the regime where the bonds are not unequal ($J_{1} \neq
J_{2}$). We consider only the antiferromagnetic region where $J_{1},J_{2} > 0$.
In the limit where $J_{2} = 0$, the system is a pure decorated square lattice
with $q_c = 2.618$ and a disordered ground state for $q = 3$. In the limit
where $J_{1} = 0$, the system is a $q = 3$ Potts model on the square lattice
formed by the decorating $B$ and $C$ sites, while the $A$ sites become
isolated. For the antiferromagnetic Potts model on the square lattice, the
critical value is $q_{c} = 3$ \cite{sokal_q_3}, and hence $T_c = 0$ here.
The phase diagram we calculate is presented in \Fig{phase}, with the critical
temperature shown in units of $J_{1}$ when $J_1 > J_2$ and $J_{2}$ when $J_2
 > J_1$.

Considering first the situation $J_2 < J_1$, we find a finite-temperature
phase transition to a state of partial order for all $J_2$, and that $T_c
\rightarrow 0$ continuously as $J_2 \rightarrow 0$. This implies that
$q_c \ge 3$ in this limit. However, it is known that $q_{c} = \frac{1}{2}(3
 + \sqrt{5}) = 2.618$ at $J_2  = 0$ \cite{wu_2,Yosio,Hajdukovic}, meaning a
disordered ground state at $T = 0$, and hence we find discontinuous behavior
here. The key property is a qualitative one, the connectivity of the lattice,
and not a quantitative one related to the coupling ratio. This result may in
fact be anticipated from the fact that the ground state should be the same
in the whole region $J_{1},J_{2} > 0$, because it corresponds to the same
coloring problem, and hence $T_c$ should be finite for all $J_2 > 0$.

We have not been able to find a similar example of such a discontinuity
anywhere in the literature. We note in this context that the applied field
acts to induce order at finite temperatures in the antiferromagnetic Ising
model on the triangular lattice and the $q = 3$ Potts model on square
lattice \cite{queiroz}, both of which are critical at $T = 0$ in the
absence of a field. In principle our method is susceptible to this type
of physics, because our calculations use a small applied field, and so we
have verified our results (\Fig{phase}) by varying $h$. Furthermore, our
system has the essential difference that the ground state of the decorated
square lattice is not critical but is fully disordered (exponentially
decaying correlations) in the absence of a field. At the other side of the
phase diagram in \Fig{phase}, obtained by taking $J_1 \rightarrow 0$, the
model becomes critical at zero temperature and fluctuations become very
large \cite{Racz}. In \Fig{phase} we show the critical temperature calculated
for moderate ratios $J_2/J_1$, and we appeal to known results in expecting
that $T_{c}$ continues smoothly to zero when $J_1 \rightarrow 0$.

We conclude our analysis of the generalized decorated square lattice
with the $q = 4$ case, where we consider only $J_{2} = J_{1}$. For this
value of $q$, the antiferromagnetic Potts model on the Union-Jack lattice
was found to have a finite-temperature phase transition to a partially
ordered ground state \cite{qiaoni,youjin}. The generalized decorated
square lattice with $J_{2} = J_{1}$ may be considered as the result of
eliminating half of the fourfold-coordinated centering sites within the
squares of the Union-Jack lattice. This weakens the constraints on the
ground state, which thus has sufficient freedom that a disordered state
may well be expected: there are four states per site and the lattice is
tripartite, with many ways to divide the system into three sublattices.

\begin{figure}[tbp]
\begin{center}
\includegraphics[scale=0.4]{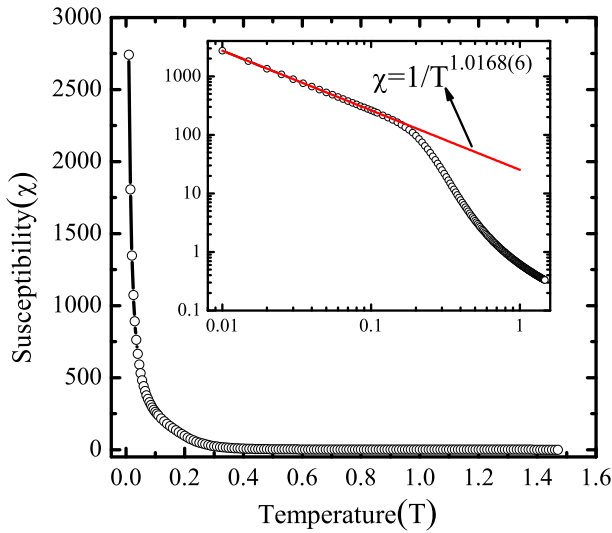}
\caption{Susceptibility for the $q = 4$ Potts model on the generalized
decorated square lattice with $J_1 = J_2$ and $D = 30$. The inset shows
the behavior near $T = 0$ on logarithmic axes, which displays good linear
scaling with a slope of $1.0168(5)$. }
\label{Mag_q-4}
\end{center}
\end{figure}

We compute the magnetization by the HOTRG method, finding no order over
the whole temperature range. However, when we calculate the susceptibility
by adding a very small magnetic field term to the system ($h/J_1 = 10^{-6}$),
we find (\Fig{Mag_q-4}) that it diverges in the limit of zero temperature.
To analyze this divergence, in the inset of \Fig{Mag_q-4} we show the
low-temperature behavior of $\chi(T)$ on logarithmic axes. The data show
a robust scaling relation of the form
\begin{equation}
\chi \sim  T^{-\gamma},
\end{equation}
where the exponent $\gamma = 1.0168(5)$. This scaling behavior indicates
that the model is critical at zero temperature. Such zero-temperature
critical behavior is also present in the antiferromagnetic $q = 2$ (Ising)
\cite{Wannier,john} and $q = 4$ Potts models on the triangular lattice
\cite{Cristopher}, as well as the $q = 3$ model on the square lattice
\cite{sokal_q_3}. From our result we can infer that the critical point
for the generalized decorated square lattice is not less than $4$, {\it
i.e.}~$q_{c} \ge 4$. In comparison with the triangular lattice, where
$q_c = 4$ \cite{Cristopher} and the coordination $z = 6$, the lattice we
consider here has a lower average coordination number, ${\bar z} = 16/3$,
but is also irregular in the sense of having inequivalent sites, an effect
now understood \cite{qiaoni} to increase $q_c$.

To conclude, we study the antiferromagnetic Potts model on the generalized
decorated square lattice by calculating its field-dependent thermodynamic
properties using the HOTRG method. For $q = 3$, we find a second-order phase
transition, occurring at $T_c = 0.528(1)$ for the isotropic point ($J_1 =
J_2$), to a state of partial order. The ground state has partial order on a
single sublattice, with spontaneous breaking of $\pi/2$ rotational symmetry
and the permutation symmetry of the Potts model. The entropy of this highly
degenerate state scales with the linear dimension of the system, meaning
that the entropy per site vanishes at zero temperature. We obtain the phase
diagram as a function of the ratio $J_{2}/J_{1}$, and our results indicate
that a finite-temperature phase transition is always present, but that
$T_c \rightarrow 0$ continuously as $J_1 \rightarrow 0$ and discontinuously
as $J_2 \rightarrow 0$. Thus the physics is determined not by the ratio
$J_2/J_1$ but by the lattice connectivity. The same model with $q = 4$ is
disordered at any finite temperature. However, for $J_1 = J_2$ we find that
the model is critical at zero temperature, the susceptibility diverging with
the form $\chi \sim  T^{-1}$, and thus we infer that the critical value of
$q$ in this lattice is $q_{c} \geq 4$.

{\it Acknowledgements:} this project was supported by the National Natural
Science Foundation of China (Grant No.~10934008, 10874215 and 11174365), the National Basic
Research Program of China (Grant No.~2012CB921704 and 2011CB309703).


\begin{thebibliography}{10}
\bibitem{wu_1} F. Y. Wu 1982 {\it Rev. Mod. Phys.}  {\bf  54} 235
\bibitem{sokal_qc} J. Salas and A. D. Sokal 1997 {\it J. Stat. Phys. }
{\bf  86} 551
\bibitem{Sokal_diced} R. Kotecky, J. Salas, and A. D. Sokal 2008 {\it Phys.
Rev. Lett.} {\bf  101} 030601
\bibitem{qiaoni} Q. N. Chen, M. P. Qin, J. Chen, Z. C. Wei, H. H. Zhao,
B. Normand, and T. Xiang 2011 {\it Phys. Rev. Lett. } {\bf  107} 165701
\bibitem{youjin} Y. J. Deng, Y. Huang, J. L. Jacobsen, J. Salas, and
A. D. Sokal 2011  {\it Phys. Rev. Lett. } {\bf  107} 150601
\bibitem{Diep_1} H. T. Diep and M. Debauche 1991 {\it Phys. Rev. } B
{\bf 43} 8759
\bibitem{Lipowski} A. Lipowski and T. Horiguchi 1995 {\it J. Phys.} A
{\bf 28} 3371
\bibitem{Diep_2} R. Quartu and H. T. Diep 1997 {\it Phys. Rev. } B {\bf 55}
2975
\bibitem{wu_2} F. Y. Wu 1980 {\it J. Stat. Phys. }  {\bf  23} 773
\bibitem{Yosio} Y. Matsuda, Y. Kasai, and I. Syozi 1981 {\it Prog. Theor.
Phys. }  {\bf  65} 1091
\bibitem{Hajdukovic} D. Hajdukovic 1983 {\it J.Phys.} A  {\bf  16} 2881
\bibitem{Nightingale} M. P. Nightingale and M. Schick 1982 {\it J. Phys. } A
{\bf  15} L39
\bibitem{den} M. P. M. den Nijs, M. P. Nightingale, and M. Schick 1982
{\it Phys. Rev. } B  {\bf  26} 2490
\bibitem{Racz} Z. Racz and T. Vicsek 1983 {\it Phys. Rev. } B  {\bf  27} 2992
\bibitem{Reed} P. Reed 1985 {\it J. Phys. } C  {\bf  18} L901
\bibitem{sokal_q_3} S. J. Ferreira and A. D. Sokal 1999 {\it J. Stat. Phys.}
{\bf  96} 461
\bibitem{Diep_4} M. Debauche, H. T. Diep, P. Azaria, and H. Giacomini 1991
{\it Phys. Rev. } B {\bf  44} 2369
\bibitem{Diep_3} P. Azaria, H. T. Diep, and H. Giacomini 1987 {\it Phys. Rev.
Lett. } {\bf  59} 1629
\bibitem{su} Y. Zhao, W. Li, B. Xi, Z. Zhang, X. Yan, S. J. Ran, T. Liu, and
G. Su 2013 {\it Phys. Rev. } E {\bf  87} 0.32151
\bibitem{levin} M. Levin and C. P. Nave 2007 {\it Phys. Rev. Lett.} {\bf 90}
120601
\bibitem{huihai} H. H. Zhao, Z. Y. Xie, Q. N. Chen, Z. C. Wei, J. W. Cai,
and T. Xiang 2010 {\it Phys. Rev. }B  {\bf  81} 174411
\bibitem{Grest} G. S. Grest and J. R. Banavar 1981 {\it Phys. Rev. Lett. }
{\bf  46} 1458
\bibitem{zhiyuan_ho} Z. Y. Xie, J. Chen, M. P. Qin, J. W. Zhu, L. P. Yang,
and T. Xiang 2012 {\it Phys. Rev. } B  {\bf  86} 045139
\bibitem{zhiyuan_srg} Z. Y. Xie, H. C. Jiang, Q. N. Chen, Z. Y. Weng, and
T. Xiang 2009 {\it Phys. Rev. Lett. } {\bf  103} 160601
\bibitem{vidal_1} G. Vidal 2007 {\it Phys. Rev. Lett. } {\bf  98} 070201
\bibitem{vidal_2} G. Vidal 2007 {\it Phys. Rev. } B  {\bf  78} 155117
\bibitem{queiroz} S. L. A. de Queiroz, T. Paiva, J. S. de Sa Martins, and
R. R. dos Santos 1999 {\it Phys. Rev. } E  {\bf  59} 2772
\bibitem{Wannier} G. H. Wannier 1950 {\it Phys. Rev }   {\bf  79} 357
\bibitem{john} J. Stephenson 1970 {\it J. Math. Phys. }  {\bf  11} 413
\bibitem{Cristopher} C. Moore and M. E. J. Newman 2000 {\it J. Stat. Phys. }
{\bf 99} 629

\end{thebibliography}
\end{document}